\documentclass[aps,pre,epsf,twocolumn,showpacs]{revtex4}
\usepackage{amsmath}
\usepackage{epsfig}
\usepackage{amssymb}

\begin{document}

\title{Interfacial adsorption in two-dimensional pure and random-bond Potts models}

\author{Nikolaos G. Fytas$^{1}$, Panagiotis E. Theodorakis$^2$, and Anastasios Malakis$^{1,3}$}

\affiliation{$^1$Applied Mathematics Research Centre, Coventry
University, Coventry CV1 5FB, United Kingdom}

\affiliation{$^2$Institute of Physics, Polish Academy of Sciences,
Al. Lotnik\'ow 32/46, 02-668, Warsaw, Poland}

\affiliation{$^3$Department of Physics, Section of Solid State
Physics, University of Athens, Panepistimiopolis, GR 15784
Zografou, Greece}

\date{\today}

\begin{abstract}
We study using Monte Carlo simulations the finite-size scaling
behavior of the interfacial adsorption of the two-dimensional
square-lattice $q$-states Potts model. We consider the pure and
random-bond versions of the Potts model for $q = 3,4,5,8$ and $q =
10$, thus probing the interfacial properties at the originally
continuous, weak, and strong first-order phase transitions. For
the pure systems our results support the early scaling predictions
for the size dependence of the interfacial adsorption at both
first- and second-order phase transitions. For the disordered
systems, the interfacial adsorption at the (disordered induced)
continuous transitions is discussed, applying standard scaling
arguments and invoking findings for bulk critical properties. The
self-averaging properties of the interfacial adsorption are also
analyzed by studying the infinite limit-size extrapolation of
properly defined signal-to-noise ratios.
\end{abstract}


\maketitle

\section{Introduction}
\label{introduction}

Critical interfacial phenomena have been studied extensively over
the last decades, both experimentally and
theoretically~\cite{Abra,Diet,Bonn,Ral}. A well-known example is
wetting, where the macroscopically thick phase, e.g., the fluid,
is formed between the substrate and the other phase, say, the gas.
Liquid and gas are separated by the interface. An interesting
complication arises when one considers the possibility of more
than two phases. A third phase may be formed at the interface
between the two other phases. An experimental realization is the
two-component fluid system in equilibrium with its vapor
phase~\cite{Diet,Mold}. Both of the above scenarios may be
mimicked in statistical physics in a simplified fashion, by either
the two-state Ising model in wetting - with the state ``+1''
representing, say, the fluid, and ``-1'' the gas - or for the case
of a third phase via multi-state spin models, simply by fixing
distinct boundary states at the opposite sides of the system. In
this latter case, the formation of the third phase with an excess
of the non-boundary states has been called as interfacial
adsorption~\cite{Huse,Fish}.

Throughout the years, various aspects of the interfacial
adsorption have been investigated via Monte Carlo methods and
density renormalization-group calculations on the basis of
specific multi-state spin models, namely Potts and Blume-Capel
models~\cite{Huse,Yeo,Kroll,Yama91,Yama,Carlon,Alba,Fytas,Alba2,Alba3}.
Additional scaling and analytic arguments have been
presented~\cite{Huse,Kroll,Carlon,Lebo,Messa,Cardy,Delf}, though
not all of them have been concretely confirmed numerically, due to
the restricted system sizes studied and the apparent underlying
scaling corrections (in some cases also because of the
uncertainties in the location of critical points). However,
notable results in the field include the determination of critical
exponents and scaling properties of the temperature and
lattice-size dependencies, as well as the clarification of the
fundamental role of the type of the bulk transition, with
isotropic scaling holding at continuous and tricritical bulk
transitions, and anisotropic scaling at bulk transitions of
first-order type.

More recently, the role of randomness on the interfacial
properties has been studied~\cite{fytas_malakis} and was found to
affect, especially, the position of the interface, the excess or
interfacial adsorption, and the form of the histograms resulting
from the different random realizations. Still, predictions of the
isotropic finite-size scaling description for the interfacial
adsorption at continuous phase transitions were observed to hold,
at least for the particular case of the dilute $8$-states Potts
model studied in Ref.~\cite{fytas_malakis}. Attention should be
drawn to related previous work on interfacial phenomena in dilute
ferromagnetic Potts models, in particular, considering
hierarchical lattices, \textit{i.e.}, applying the Migdal-Kadanoff
real space renormalization to the square lattice~\cite{Monthus},
or performing a preliminary Monte Carlo study for the square
lattice model~\cite{Brener}.

Motivated by Ref.~\cite{fytas_malakis}, in the present work we
study the scaling behavior of the interfacial adsorption of
several two-dimensional pure and random-bond Potts models. In
particular we consider the disordered $q = 3$ and pure $q = 8$
models, that complement our previous work~\cite{fytas_malakis},
and we furthermore extend these studies by presenting new results
for both the pure and disordered versions of the $q = 4$, $5$, and
$q = 10$ models. For the case of pure and randomness-induced
continuous transitions we present concrete numerical evidence in
favor of the standard isotropic scaling with exponents that can be
traced back to the best-known estimates of the bulk critical
exponent ratio $\beta/\nu$ of the Potts model, where $\beta$ and
$\nu$ are the bulk critical exponents of the order parameter and
correlation length, respectively, thus reinforcing the main result
of Ref.~\cite{fytas_malakis} for the $q=8$ case. For the
first-order phase transitions corresponding to the pure $q=5$,
$8$, and $q = 10$ Potts models, our numerical data and scaling
analysis strongly support the early scaling predictions for the
size dependence of the interfacial adsorption at first-order
transitions~\cite{Kroll}. In the present paper we also discuss the
self-averaging properties of the interfacial adsorption of the
disordered Potts models in terms of properly defined
signal-to-noise ratios, an aspect that hasn't been considered
before in the literature. Various forms of corrections-to-scaling
are discussed and depending on the number of states $q$ of the
Potts model some expectations from the literature are used as the
best possible choices. However, the overall observed scaling
behavior does not drastically change by the use of such
corrections.

The outline of the article is as follows: In Sec.~\ref{model} the
model and the interfacial adsorption are introduced and in
Sec.~\ref{simulation} the numerical method implemented is
outlined. Our main finite-size scaling analysis and results are
presented in Sec.~\ref{results}. The summary,
Sec.~\ref{conclusions}, concludes the article.

\section{Model and interfacial adsorption}
\label{model}

We study the nearest-neighbor $q$-states Potts model on the square
lattice described by the Hamiltonian
\begin{equation}
\label{eq:Hamiltonian} \mathcal{H} = -\sum_{\langle ij
\rangle}J_{ij}\delta_{\sigma_i,\sigma_j}.
\end{equation}
The Potts variable at site $i$, $\sigma_{i}$, takes the values
$1,2,\ldots,q$~\cite{Wu} and the ferromagnetic random couplings
$J_{ij}>0$ between nearest-neighbor sites $i$ and $j$ are either
$J_1$, with probability $p$, or $J_2$, with probability $1-p$. In
the case $J_1 > J_2$, one has either strong or weak bonds. Then,
the ratio $r=J_2/J_1$ defines the disorder strength. Clearly, the
value $r=1$ corresponds to the pure model.

In this article, we shall consider the system at its self-dual
point, where both couplings occur with the same probability,
$p=1/2$. Then, the phase-transition temperatures between the
ordered ferromagnetic phase and the high-temperature disordered
phase are known exactly from self-duality for arbitrary values of
the internal states $q$ and disorder-strength ratios
$r$~\cite{Kinzel}
\begin{equation}
\label{eq:Critical point} \left(e^{(J_1/k_{\rm B} T_{\rm
c})}-1\right) \left(e^{(rJ_1/k_{\rm B}T_{\rm c})}-1 \right) = q.
\end{equation}
From the above equation one may easily numerically calculate
$k_{\rm B}T_{\rm c}/J_1$ for any given value of $r$ ($r=1/10$ at
the present study). Thus, via Eq.~(\ref{eq:Critical point}),
analyzes on the critical behavior of the interfacial adsorption of
the disordered Potts model, based on Monte Carlo simulation data,
as it is also done in the present paper, are significantly
simplified.

\begin{figure}
  \begin{center}
    \includegraphics*[width=7 cm]{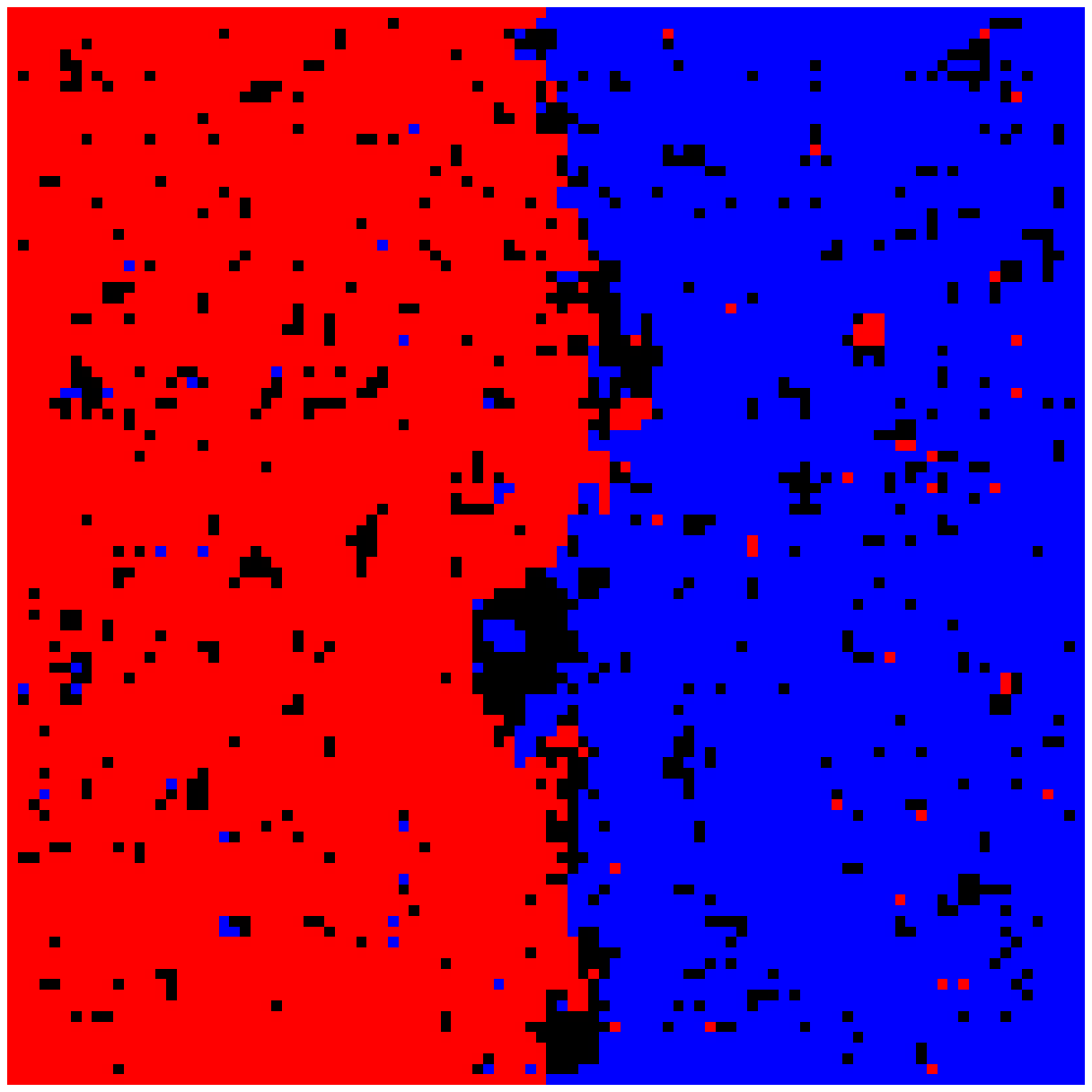}
    \includegraphics*[width=7 cm]{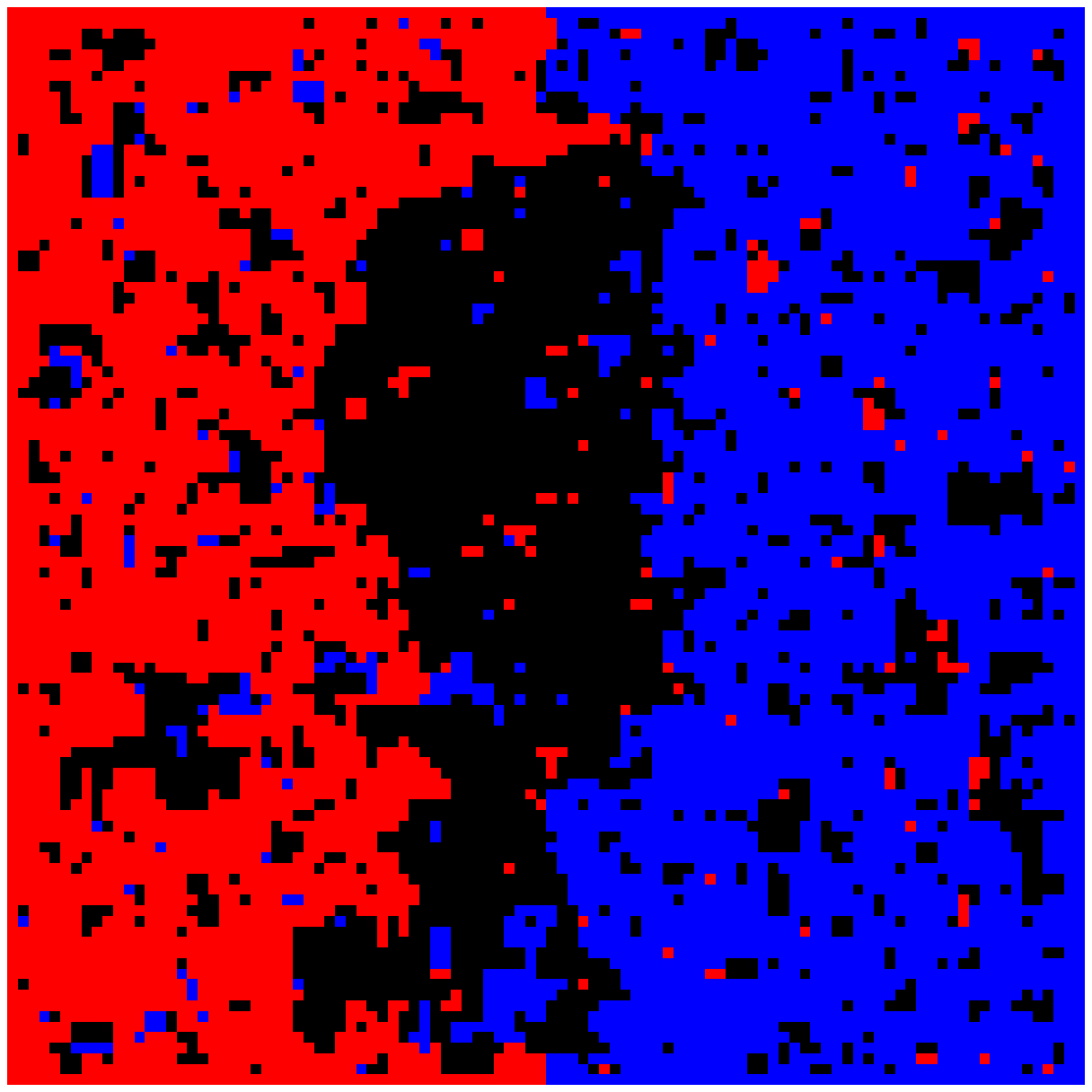}
  \end{center}
  \caption{\label{fig_conf_random_q10}
    (color online) Typical equilibrium Monte Carlo configurations of
an $L = 100$, $q = 10$, pure (upper panel) and random-bond
$r=1/10$ (lower panel) Potts model at $k_{\rm B} T/J_1 = 0.98
k_{\rm B} T_{\rm c}/J_1$. In both cases, red color depicts the
$q=1$ states, blue color the $q = 2$ states, whereas the
non-boundary states ($q \ge 3$) adsorbed at the interface are
shown blackened. Note that the fixed boundaries $[1:2]$ are also
included in these illustrations.
  }
\end{figure}

Bulk criticality of such disordered Potts models on the square
lattice has attracted much interest, partly, because the
transition is of continuous type for all values of $q$, while
being, in the pure case, of first order for
$q>4$~\cite{Wu,Berche}. Exact values of the critical exponents are
only known in the clean case~\cite{Wu}. Numerical analysis, in the
dilute case suggest that the bulk critical exponents depend rather
mildly on $q$~\cite{Berche}. Then, the analysis of the interfacial
adsorption in these models may be simplified by the fact, that
isotropic finite-size scaling is expected to hold at continuous
transitions~\cite{Huse,Kroll,Yama91}. Static and dynamic bulk
critical properties of the disordered Potts models have been
estimated, using a variety of, predominantly, numerical
methods~\cite{Berche}.

The degeneracy between the $q$ equivalent Potts states may be
lifted by appropriate boundary conditions. In order to study the
interfacial adsorption, denoted hereafter as $W$, we shall employ
special boundary conditions, distinguishing the cases $[1:1]$ and
$[1:2]$. For the case $[1:1]$, the Potts variable is set, at all
boundary sites, equal to $q=1$, while for the case $[1:2]$, the
variable is set equal to $1$ at one half of the boundary sites,
and to $2$ at the opposite half of the boundary sites. Then, the
boundary condition $[1:2]$ introduces an interface between the
1-rich domain (or phase) and the 2-rich domain (or phase). By
examining typical Monte Carlo equilibrium configurations, as shown
in Fig.~\ref{fig_conf_random_q10} for an instance of the pure
(upper panel) and disordered (lower panel) $q=10$ Potts model, it
is seen that at the interface between the 1- and and 2-rich
domains an excess of the non-boundary states is generated compared
to the case of the absence of an interface. As expected, in the
dilute case, the position of the interface, as well as the extent
of the intervening third phase of non-boundary states, may be
strongly affected by the spatial distribution of the couplings.

Then, the interfacial adsorption measuring the surplus of
non-boundary states induced by the interface between the 1- and
2-rich regions for lattices with $L^{2}$ non-boundary sites,
where $L$ denotes the linear dimension of the lattice, is defined
following Eq. (2) of Ref.~\cite{Carlon} by
\begin{equation}
\label{eq:ia} W=\frac{1}{L} \sum_n \sum_{i}
\left(\langle\delta_{\sigma_{i},n}\rangle_{[1:2]}-\langle
\delta_{\sigma_{i},n}\rangle_{[1:1]}\right).
\end{equation}
The summation is over all non-boundary sites $i$ and all non-boundary
states $n=3,4,\ldots,q$ and the
thermal average is taken. For a disordered system now, the above
definition denotes the interfacial adsorption of a single
realization. In this case, a second
average over the disorder distribution needs to be taken, as
it is also done in the present work.
The above definition of the interfacial
adsorption has been successfully used in the
past~\cite{Huse,Yeo,Kroll,Yama91,Yama,Carlon} and although its
implementation demands Monte Carlo simulations on two systems with
different boundary conditions, it offers a simple and physically
appealing method to estimate the width of the interface by
reflecting quantitatively the difference in fluctuations of the
two systems. Thus, $W$ is geometrically interpreted as the
effective width of the domain of non-boundary states between the
1- and 2-rich domains. In the case of second-order phase
transitions, normalizing Eq.~(\ref{eq:ia}) with $1/L$ produces an
effective width that gives divergencies of the form $\sim
L^{(1-\beta/\nu)}$~\cite{Huse,Yeo,Kroll}. However, one may also
argue that it makes more sense to normalize Eq.~(\ref{eq:ia}) with
$1/L^{2}$. In this case the corresponding effective width scales
as $\sim L^{-\beta/\nu}$ with the system size and we should
recognize that in this practise the effective width becomes of
zero measure compared to the system's size.

\section{Simulation details}
\label{simulation}

In our simulations of the Potts models we applied the Metropolis~\cite{metropolis}
and the cluster-flip Wolff algorithm~\cite{wolff}. Of course,
cluster flips violating the boundary conditions are not
allowed~\cite{Gamsa}. As usual, small lattices may be simulated
using the Metropolis algorithm, while the Wolff algorithm~\cite{wolff} is more
efficient and is preferred for larger, say $L > 30$, system sizes~\cite{LanBin}.
Overall, we studied lattices with up to $100\times 100$ sites for
both pure and disordered Potts models. Only for the pure $q=3$
Potts model data for system sizes up to $200^2$ sites have been
generated and taken from Ref.~\cite{fytas_malakis}.

Certainly, equilibration and averaging times depend on the lattice
size. Moreover, for disordered models, we observed that the given
bond realization may affect these times. In the case of the
Metropolis algorithm, eventually, simulations with $10^7$ Monte
Carlo steps per site for $L=10$ were performed, increasing the
length of the runs, roughly, with $L^2$. On the other hand, for
the application of the Wolff algorithm, the number of (Wolff)
clusters used in our simulations varied from $2\times 10^7$ for
the smaller systems sizes up to $3\times 10^9$ for the larger
sizes considered. The Wolff clusters are constructed as usual with
the appropriate acceptance probability from the set of the
neighboring lattice sites sharing the same value of the spin~\cite{wolff}.

In disordered systems, the main source of errors stems from the
fact that the simulation data may vary drastically among different
random configurations. In this work, the corresponding histograms
or distributions have been recorded for $r=1/10$ and all values of
$q$ considered. Noteworthy, bulk properties of the random-bond
Potts model for various values of $r$ and $q$ have been studied
quite extensively
before~\cite{Berche,Chen,Jacob,picco96,Chat,Chat99,Pala,Picco} and
the obtained pool of results will prove to be extremely useful for
the analysis in the following Section. The histograms at the
critical point show nearly Gaussian shapes, but being weakly
tailed, in accordance with previous observations and discussions
in our previous work~\cite{fytas_malakis} and in
Ref.~\cite{Monthus} for the dilute Potts models on hierarchical
lattices. The standard errors resulting from an ensemble average
over bond realizations decrease with the number of configurations,
$\mathcal{N}$, as $\sim 1/\sqrt{\mathcal{N}}$. The proportionality
factor seems to become somewhat smaller for larger lattices. To
obtain reasonable accuracy, we averaged over a large number of
different bond configurations, varying from $\mathcal{N} =
20\times 10^3$ for the smaller system sizes studied down to $10^3$
for the larger ones. For pure Potts models ($r=1$) error bars
follow from averaging over a few hundreds of Monte Carlo runs
employing different random numbers, as usual. Finally, for the
application of finite-size scaling on the numerical data in terms
of characteristic power-law fittings as will be discussed below,
we employed the standard criterion of the $\chi^{2}/{\rm DOF}$,
where DOF denotes the number of degrees of freedom.

\section{Finite-size scaling analysis}
\label{results}

To determine critical properties from Monte Carlo data we use
finite-size scaling arguments. For the interfacial adsorption,
$W$, one expects~\cite{Huse}
\begin{equation}
\label{eq:scal} W \approx  L^{a}\Omega(t L^{1/\nu}),
\end{equation}
where the critical exponent $a$ is determined as mentioned above
by the bulk critical exponents $\beta$ and $\nu$ via~\cite{Huse}
\begin{equation}
\label{eq:ab} a = 1- \frac{\beta}{\nu},
\end{equation}
$t=|T-T_{\rm c}|/T_{\rm c}$ is the reduced critical temperature
and $\Omega$ the scaling function. A more refined ansatz invokes
corrections to the asymptotic scaling behavior, as will be
discussed below. Note however that in the present work, we are not
interested in the temperature dependence of $W$, but rather only
on its size dependence. From the above scaling
assumption~(\ref{eq:scal}) we derive that the leading critical
behavior of the interfacial adsorption is given by $W \sim
L^{a}$~\cite{Huse}.

At the critical (and tricritical) points, the singularities in the
interfacial adsorption are induced by bulk critical fluctuations.
On the other hand, at first-order transitions there are no bulk
critical fluctuations and the divergence of $W$ arises from an
interface delocalisation transition~\cite{Lipo}. In the latter
case, for lattices of square shapes, a linear divergence of the
form $W \sim L$ is expected for the interfacial adsorption at the
transition point according to the arguments of Selke \emph{et
al.}~\cite{Kroll}.

The above scaling predictions have been confirmed reasonably well
in previous Monte Carlo simulations for pure $q=3$ Potts and
Blume-Capel models showing continuous
transitions~\cite{Huse,Yeo,Kroll,Yama91}. On the other hand for
the case of first-order transitions where metastability effects
cast further difficulties in the scaling analysis, the expected
linear divergence has been only partially supported by the
numerical data at hand~\cite{Kroll,Yama}. In the present study, we
extend and refine previous results on the size-dependence of the
interfacial adsorption by considering the pure Potts model for
various values of the internal states $q$ as well as by including
bond randomness in the system.

We start our presentation of results by contrasting the critical
interfacial adsorption of the pure and disordered 3-states Potts
model in Fig.~\ref{fig_W_q3} (the underscore symbols p(r) in
Fig.~\ref{fig_W_q3}, and following figures as well, refer to the
pure(random) cases, respectively). The data for the pure model
(filled circles) were taken from Ref.~\cite{fytas_malakis} and
have been analyzed according to that paper by fitting them (black
line in Fig.~\ref{fig_W_q3}) to equation
\begin{equation}
\label{eq:iaL_cor} W \sim L^{a} (1+b L^{-x}),
\end{equation}
taking into account a possible leading corrections-to-scaling
exponent $x$. Fixing $x$ to the expected value
$4/5$~\cite{DoFa,Priv,Shchur,Queiroz} the resulting estimate for
the critical exponent $a_{\rm p} = 0.870(3)$ agrees within errors
with the predicted exact value $a=13/15=0.866\cdots$ (we remind
the reader that $(\beta / \nu)_{\rm p} = 2/15 =
0.133\cdots$)~\cite{Wu}. Accordingly, the findings on the pure
3-states Potts model strongly support the correctness of the
finite-size scaling description~(\ref{eq:scal}).
\begin{figure}[htbp]
\includegraphics*[width=8 cm]{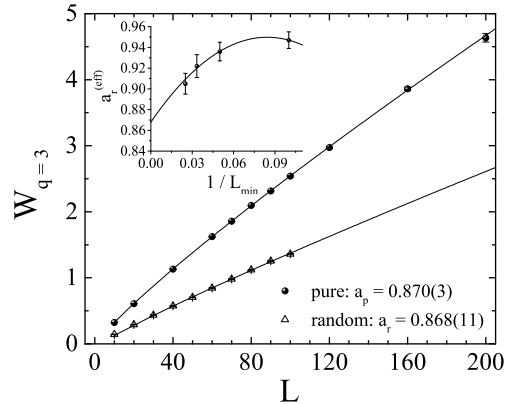} \caption{\label{fig_W_q3}
Finite-size scaling of the interfacial adsorption of the pure and
random-bond $q=3$ Potts model. The inset illustrates the infinite
limit-size extrapolation of the effective exponent $a^{\rm
(eff)}_{\rm r}$.}
\end{figure}
However, we point out that the influence of the
corrections-to-scaling exponent $x$ is only marginal in these
estimations, since, using $x = 1$ we find $a_{\rm p} = 0.873(4)$
which a bit larger than the expected result, whereas using $x =
3/5$ we find $a_{\rm p} = 0.867(2)$, which is closer to the exact
value.

We continue our presentation with the disordered $q = 3$ Potts
model. Following previous considerations on its bulk critical
properties, we set $r = 1/10$, where the randomness dominated
behavior is expected to show up already for moderate lattice
sizes. The arguments leading to this observation were originally
discussed by by Wang \emph{et al.}~\cite{wang} for the dilute
Ising model and later used for other Potts model as
well~\cite{Berche,Chen,Jacob,picco96,Chat,Chat99,Pala,Picco}. In
particular, we monitored, in our simulations, the size dependence
of the critical interfacial adsorption. Numerical results for the
$q=3$ disordered model are depicted by the open triangles in
Fig.~\ref{fig_W_q3}. Since for the present case and for the
subsequent random cases (apart from the $q = 4$ case for which
logarithmic corrections are known to exist and will be taken into
account) there is no clear information with respect to the leading
corrections-to-scaling in the literature, and in the light of the
above discussion for a marginal effect of the
corrections-to-scaling exponent in the fits, we will use the value
$x = 1$ in Eq.~(\ref{eq:iaL_cor}) which is the simplest choice
leading to reasonable fits. In this way we obtain effective
exponents by varying the lower system size $L_{\rm min}$ included
in the fits. A second-order polynomial extrapolation of these
effective exponents, as illustrated in the corresponding inset of
Fig.~\ref{fig_W_q3}, provides us with the value $a_{\rm r} =
0.868(11)$. This estimate of $a_{\rm r}$ is compatible to the
value $0.8679(3)$, if one accepts Eq.~(\ref{eq:ab}) and the
estimate $(\beta/\nu)_{\rm r} = 0.1321(3)$ for the bulk critical
exponent ratio of the dilute $q=3$  Potts model given in
Ref.~\cite{Chat99}.
\begin{figure}[htbp]
\includegraphics*[width=8 cm]{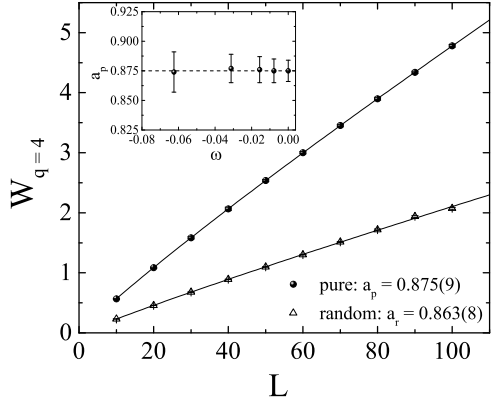} \caption{\label{fig_W_q4}
Finite-size scaling of the interfacial adsorption of the pure and
random-bond $q=4$ Potts model. The inset shows fitting results for
the exponent $a_{\rm p}$ of the pure system by varying the value
of the corrections exponent $\omega$ within the regime
$[-1/16,0]$. The dashed line marks the value $a_{\rm p} =
0.875(9)$ that corresponds to the case $\omega = 0$.}
\end{figure}
Thus, we may conclude that for the disordered $q=3$ case as well
$a_{\rm r} = 1 - (\beta/\nu)_{\rm r}$, in accordance with the
finite-size scaling ansatz~(\ref{eq:scal}), although it is true
that for this particular case there seems to be hardly any
difference among the exact value of the exponent ratio $\beta/\nu$
of the pure model and the corresponding estimates for the
disordered model. This fact has also been underlined in an
extensive study of the magnetization of the $q=3$ random-bond
Potts model by Picco~\cite{picco96}.

Next, we consider the delicate $q=4$ model, for which the size
dependence of the interfacial adsorption has never been studied
previously. As it is well-known, this is a borderline case of the
Potts universality class, where logarithmic corrections are known
to exist. In particular, Salas and Sokal~\cite{salas} have studied
in detail the form of these scaling corrections and have found
multiplicative logarithmic corrections as well as additive
logarithmic corrections, some of which are universal.
Of course, numerically observing the existence of logarithmic
corrections is always difficult, and it is almost impossible to
detect logarithmic corrections using system sizes of the order of
$L = 100$, as in the present work. However, given the description
of these logarithmic corrections in Eqs. (3.21) and (3.22) in
Ref.~\cite{salas}, we may try to fit our numerical data using this
prescription. Accordingly, a plausible finite-size scaling ansatz
for the $q = 4$ Potts model has the general form~\cite{salas}
\begin{equation} \label{eq:log}
Q \sim L^{a} [\ln{(L)}]^{\omega}
\left(1+b\frac{\ln{[\ln{(L)}]}}{\ln{(L)}}+b'\frac{1}{\ln{(L)}}\right).
\end{equation}
The corrections exponent $\omega$ takes the value $-1/16$ for the
absolute magnetization, whereas $\omega = -1/8$ for the magnetic
susceptibility and $\omega = -3/2$ for the specific
heat~\cite{salas,caselle99,lev09}. In the light of the previous
expectations~\cite{Huse}, and given that $W$ scales with an
exponent $\beta/\nu$ [see Eq.~(\ref{eq:ab}], which also describes
the finite-size scaling behavior of the absolute magnetization,
one may be tempted to use an exponent $\omega = -1/16$.
\begin{figure}[htbp]
\includegraphics*[width=8 cm]{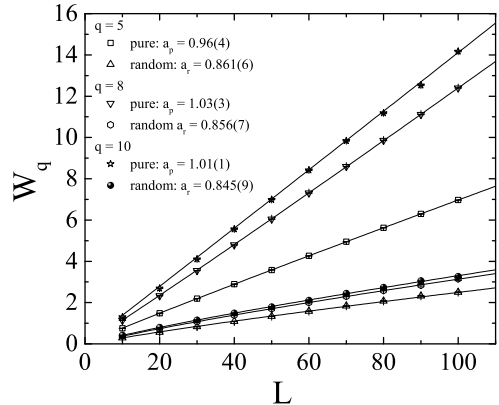} \caption{\label{fig_W_first}
Finite-size scaling of the interfacial adsorption of the pure and
random-bond Potts model for various values of the internal states
$q$ in the originally first-order regime, as indicated.}
\end{figure}
However, the term $[\ln{(L)}]^{-1/16}$ for the system sizes we
studied in the present work is close to $1$ and we therefore
expect that it will have no severe effect in the following fitting
attempts. Indeed, we have performed fits of the
form~(\ref{eq:log}) on the numerical data of the pure $q = 4$
system (shown by filled circles in the main panel of the figure)
for several candidate values of $\omega$ within the regime
$[-1/16,0]$ that verify this expectation. An illustrative plot of
our analysis is presented in the inset of Fig.~\ref{fig_W_q4}
where the estimated values of the exponent $a_{\rm p}$ are plotted
as a function of the fixed exponent $\omega$ used in the fit. The
dashed line marks the value $a_{\rm p} = 0.875(9)$ that
corresponds to the case $\omega = 0$ (illustrated also by the
solid black line in the main panel of the figure with an
$\chi^2/{\rm DOF} \approx 0.9$ merit of the fit). Correspondingly,
the open triangles in Fig.~\ref{fig_W_q4} present our numerical
data for the disordered $q = 4$ Potts model, for which a fit of
the form~(\ref{eq:log}) with $\omega = 0$ gives the result $a_{\rm
r} = 0.863(8)$, as also indicated in the panel. Again, isotropic
scaling and Eq.~(\ref{eq:ab}) is satisfied, to a high accuracy for
the pure model, for which $(\beta/\nu)_{\rm p} = 1/8 = 0.125$, and
within errors for the disordered model for which $(\beta/\nu)_{\rm
r} = 0.1385(3)$~\cite{Chat99}.

The second part of our study refers to the interfacial properties
at the originally first-order transition regime of the Potts
model, \textit{i.e.}, for $q>4$. We have simulated the model with
internal states $q=5$, $8$, and $q=10$ in both its pure and
disordered ($r=1/10$) version. The data for the disordered $q=8$
case have been taken from Ref.~\cite{fytas_malakis}.
\begin{figure}[htbp]
\includegraphics*[width=8 cm]{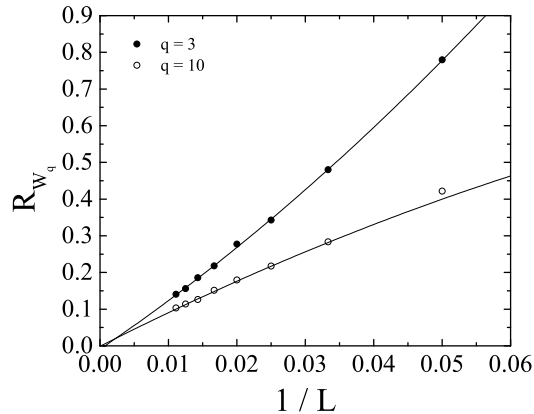} \caption{\label{fig_R_W}
$R_{\rm W}$ as a function of the inverse linear system size for
two values of $q$, as indicated. The solid lines are second-order
polynomial extrapolations to the limit $L\rightarrow \infty$.}
\end{figure}
Our numerical results for the critical interfacial adsorption and
the relevant scaling analysis are illustrated in
Fig.~\ref{fig_W_first}. Several comments are in order: (i) For the
pure system we find a clear linear divergence of $W$ for all
values of $q > 4$ as predicted by scaling arguments~\cite{Kroll},
since fittings of the form $W \sim L^{a_{\rm p}}$ give estimates
of $a_{\rm p} \approx 1$ without the need of including scaling
corrections. (ii) For the corresponding disordered system fittings
of the form~(\ref{eq:iaL_cor}) with a fixed correction of $x = -1$
give estimates of $a_{\rm r}$ that again support the isotropic
scaling and Eq.~(\ref{eq:ab}). Note the most accurate existing
estimates of the ratio $(\beta/\nu)_{\rm r}$ are $0.141(3)$,
$0.145(5)$, and $0.155(5)$ for $q=5$, $8$, and $q=10$,
respectively~\cite{Jacob,picco96,Chat,Chat99,Pala,Picco}. (iii)
Using a wide range of internal states within the regime $q = 3 -
10$ and two versions of the Potts model, namely the pure model and
its disordered counterpart, we have shown that isotropic scaling
holds, as well as the relation $a = 1 - \beta/\nu$.

For all systems studied we also recorded, in addition to the
interfacial properties, standard thermodynamic quantities, for
both types of boundary conditions. In particular, we measured the
specific heat $C_{1:1}$ and $C_{1:2}$ and the order parameter
given by the majority fraction of the Potts states~\cite{Wu},
$m_{1:1}$ and $m_{1:2}$. Fittings of $m_{1:1}$ and $m_{1:2}$,
vanishing at $T_{\rm c}$ as $\sim L^{-\beta/\nu}$, gave estimates
of the magnetic exponent ratio compatible to the above presented
results, thus giving further credit to the current numerical data.
It is also interesting to note that although the specific-heat
data for the pure $q>4$ systems could not be fully described by
the standard $L^{2}$ scaling behavior for systems with linear
sizes up to $L=100$ (possibly due to strong corrections,
especially for the case of the $q=5$ weak first-order transition),
we have been able to probe nicely the first-order character of the
transition based on the linear divergence of the interfacial
adsorption shown for system sizes $L\leq 100$.

As a large part of the current contribution is based on the
disordered version of the Potts model, we close this Section with
an illustration of the self-averaging properties of the
interfacial adsorption. As we know, our numerical studies of
disordered systems are carried out using finite samples; each
sample is a particular random realization of the quenched
disorder. A measurement of a thermodynamic property, say $W$ for
the interfacial adsorption considered here, yields a different
value for every sample. In an ensemble of disordered samples of
linear size $L$ the values of $W$ are distributed according to a
probability distribution. The behavior of this distribution is
directly related to the issue of self-averaging. In particular, by
studying the behavior of the width of this distribution, one may
address qualitatively the issue of self-averaging, as has already
been stressed by previous authors. In general, we characterize the
distribution by its disorder average $[W]$, and also by the
relative variance $R_{W}=V_{W}/[W]^{2}$, where $V_{W} =
[W^2]-[W]^2$. The limiting value of this ratio is indicative of
the self-averaging properties of the system~\cite{WD95,AH96}. In
Fig.~\ref{fig_R_W} we show the infinite limit-size extrapolation
of the ratio $R_{W}$ for the random-bond (again $r=1/10$) Potts
model and two selected values of $q$ as indicated in the figure,
namely the values $q=3$ (filled circles) and $q=10$ (open
circles). For both cases we find that $R_{W}\rightarrow 0$ as
$L\rightarrow \infty$, indicating that the interfacial adsorption
restores self-averaging in the thermodynamic limit. Similar
fitting attempts for other values of $q$, not shown here for
brevity, support our conclusion based on these data.

\section{Conclusions}
\label{conclusions}

We performed extensive Monte Carlo simulations to study the
critical interfacial properties in pure and disordered
ferromagnetic $q$-states Potts models on the square lattice for
various values of the internal states $q\in\{3 - 10\}$. Interfaces
have been introduced by fixing the Potts variables at opposite
sites in two different states. The local Metropolis and
cluster-flip Wolff algorithms have been used to simulate all
systems at their critical points, taking advantage of the existing
self-duality. For the disordered models an extensive disorder
averaging has been performed in order to control the
sample-to-sample fluctuations of the model. The finite-size
scaling analysis on our wide-range numerical data allowed us to
safely conclude that the isotropic finite-size scaling description
for the interfacial adsorption at (pure and randomness-induced)
continuous phase transitions holds. Additionally, for the pure $q
> 4$ systems that undergo a (weak: $q = 5$, or strong: $q=8$ and
$q = 10$) first-order phase transition, we have been able to probe
the linear divergence of the interfacial adsorption, verifying the
early predictions of the scaling theory. Finally, we have
discussed the self-averaging properties of the interfacial
adsorption by studying the infinite limit-size extrapolation of
properly defined signal-to-noise ratios and we have found that
self-averaging is restored in the thermodynamic limit.

\acknowledgments{The authors would like to thank Prof. W. Selke
for suggesting this topic of research, his ongoing collaboration,
as well as for many useful discussions and critical comments. We
would also like to thank an anonymous Referee for his instructive
comments on the problem of corrections-to-scaling in the Potts
model. N.G. F. is grateful to Coventry University for providing a
Research Sabbatical Fellowship during which this work has been
completed.}

{}

\end{document}